\documentclass[12pt,prb,draft]{revtex4}
\newcommand{\be}{\begin{equation}}
\newcommand{\ee}{\end{equation}}

\newcommand{\bea}{\begin{eqnarray}}
\newcommand{\eea}{\end{eqnarray}}
\newcommand{\om}{\omega}
\newcommand{\Om}{\Omega}

\newcommand{\Eq}[1]{Eq.(\ref{#1})}

\begin{document}
\title{Simplified derivation of the Hawking-Unruh temperature for an accelerated observer in vacuum}
\author{Paul M. Alsing}
\address{Department of Physics and Astronomy \\
University of New Mexico\\
Albuquerque, New Mexico 87131\\
alsing@hpc.unm.edu}
\author{Peter W. Milonni}
\address{Theoretical Division (T-DOT)\\
Los Alamos National Laboratory\\
Los Alamos, New Mexico 87545\\
pwm@lanl.gov}
\date{\today}

\begin{abstract}
A detector undergoing uniform
acceleration $a$ in a vacuum field responds just as though it were
immersed in thermal radiation of temperature $T=\hbar a/2\pi kc$.
A simple, intuitive derivation of this result is given for the
case of a scalar field in one spatial dimension.
The approach is then extended to treat
the case where the field seen by the accelerated observer
is a spin-1/2 Dirac field.
\end{abstract}
\maketitle

\section{Introduction}
Hawking \cite{hawking} predicted that a black hole should radiate with a temperature $T=\hbar g/2\pi kc$, where
$g$ is the gravitational acceleration at the surface of the black hole, $k$ is Boltzmann's constant, and $c$
is the speed of light. This results from the effect of the strong gravitation on the vacuum field.
Shortly thereafter it was shown separately by Davies and Unruh
that a uniformly accelerated detector in vacuum responds just as though it were in a
thermal field of temperature \cite{unruh,davies,also,takagi,davies2}
\be
T={\hbar a\over 2\pi kc} \ ,
\label{eq1}
\ee
where $a$ is the acceleration in the instantaneous rest frame of the detector.
These results suggest profound consequences for the merger of quantum
field theory and general relativity and sparked intense debates
over unresolved questions that are still actively investigated today:
(1) if black holes are not really "black," are naked singularities
the ultimate fate of black holes, or will a long-sought fusion of quantum mechanics
and general relativity into a coherent theory of quantum gravity prevent
such occurrences?; (2) if a quantum mechanical pure state is dropped
into a black hole and pure thermal (uncorrelated) radiation results, how does one explain
the apparent non-unitary evolution of a pure state to a mixed state?

This intriguing result of quantum field theory
has arguably not been derived in any physically intuitive way.
Numerous explicit and detailed calculations have appeared in the
scientific literature over the last, roughly 30 years for a wide
host of spacetimes. However, even for the "simplest"
calculation involving a scalar field, the intricacies of
field theory techniques, coupled with a forest of special function
properties, makes most derivations intractable for the curious, casual
scientific non-specialist. An investigation of the Unruh effect
for the case of Dirac particles of spin $1/2$ brings in a whole
host of new machinery, least of which is the formulation of
the Dirac equation in curvilinear coordinates (i.e. essentially
in curved spacetime). This quickly goes beyond the expertise of
most casual scientific readers. However, in both cases,
this beautiful and important result can be stated quite simply:
for a scalar field (bosons) the accelerated observer
sees a Bose-Einstein (BE) distribution at temperature $T$ given
by \Eq{eq1}, while for a spin-$1/2$ field the accelerated observer
sees a Fermi-Dirac distribution at the same temperature.

It is the purpose of this paper to present simplified derivations of \Eq{eq1}
in a way that is suitable for advanced undergraduate or beginning graduate students
and that elucidates the essential underlying  physics of the Unruh effect.
Once one accepts the simplest features of a quantized vacuum field, the result \Eq{eq1} emerges
as a consequence of time dependent Doppler shifts in the field seen by the accelerated observer.

In the following section the essential features of uniform acceleration for our purposes are reviewed, and in Section 3
we use these results to obtain \Eq{eq1} in an almost trivial way based on the Doppler effect. In Section 4 this
simple approach to the derivation of the temperature \Eq{eq1} is developed in more detail.
In Section 5 we extend the previous calculations for scalar fields to spin-$1/2$ Dirac fields.
We close with a brief summary and discussion.

\section{Uniform Acceleration}
Uniform acceleration is defined as a constant acceleration $a$
in an instantaneous inertial frame in which the observer is at rest. The acceleration $dv/dt$ in the lab frame is related to $a$ by
the Lorentz transformation formula
\be
{dv\over dt}=a\left(1-{v^2\over c^2}\right)^{3/2} \ .
\label{eq2}
\ee
Integrating, and taking $v=0$ at $t=0$, we have $v(t)=at/\sqrt{1+a^2t^2/c^2}$. The relation $dt=d\tau/\sqrt{1-v^2/c^2}$ between the
lab time ($t$) and the proper time ($\tau$) for the accelerated observer gives $t(\tau)=(c/a)\sinh(a\tau/c)$ if we take $t(\tau=0)=0$.
The velocity $v$ of the accelerated observer as seen from
the lab frame can be expressed in terms of the proper time $\tau$ as
\be
v(\tau)=c\tanh\left({a\tau\over c}\right) \ .
\label{eq3}
\ee
A straightforward integration of the above equations yields the well known hyperbolic orbit of the
accelerated (Rindler) observer in the $\mathbf{z}$ direction \cite{rindler}:
\be
\label{eq3.5}
 t(\tau) = \frac{c}{a} \, \sinh\left(\frac{a\tau}{c}\right), \qquad z(\tau) = \frac{c^2}{a} \, \cosh\left(\frac{a\tau}{c}\right) \ .
\ee
Throughout this work we consider $a>0$, i.e. the observer accelerates in the $\mathbf{z}$ direction.

\section{Indication of Thermal Effect of  Acceleration}
Consider now a plane-wave field of frequency $\om_K$ and wave vector ${\bf K}$ parallel or anti-parallel to the direction ${\bf z}$ along
which the observer is accelerated. In the instantaneous rest frame of the observer
the frequency $\om_K'$ of this field is given by the Lorentz transformation formula
\be
\om_K'(\tau)={\om_K-Kv(\tau)\over\sqrt{1-v^2(\tau)/c^2}}={\om_K[1-\tanh({a\tau\over c})]\over\sqrt{1-\tanh^2({a\tau\over c})}}=\om_K e^{-a\tau/c}
\ \ \ \ (K=+\om_K/c)
\label{eq4}
\ee
for $K=+\om_K/c$, i.e.,  for plane-wave propagation along the direction ${\bf z}$ of the observer's acceleration. For propagation in the
$-{\bf z}$ direction, similarly,
\be
\om_K'(\tau)=\om_K e^{a\tau/c} \ \ \ \ (K=-\om_K/c) \ .
\label{eq5}
\ee
Note that, for small values of $a\tau$, $\om_K'\cong \om_K(1\mp a\tau/c)$, the familiar Doppler
shift. Equations (\ref{eq4}) and (\ref{eq5}) involve {\it time-dependent} Doppler shifts seen by the accelerated observer.

Because of these Doppler shifts our accelerated observer sees waves with a time-dependent phase
$\varphi(\tau)\equiv \int^{\tau}\om_K'(\tau')d\tau' = (\om_Kc/a)\exp(a\tau/c)$.
We suppose therefore that, for a wave propagating in the $-{\bf z}$ direction, for which
$\int^{\tau}\om_K'(\tau')d\tau'=(\om_Kc/a)\exp(a\tau/c)$, he sees a frequency spectrum $S(\Om)$ proportional to
\be
\label{eq6}
\left|\int_{-\infty}^{\infty}d\tau e^{i\Om\tau}e^{i(\om_Kc/a)e^{a\tau/c}}\right|^2 \ .
\ee
 Changing variables to $y=e^{a\tau/c}$, we have
\bea \int_{-\infty}^{\infty}d\tau e^{i\Om\tau}e^{i(\om_K
c/a)e^{a\tau/c}}&=&{c\over a}\int_0^{\infty}dy\,y^{(i\Om c/a-1)}e^{i(\om_Kc/a)y} \nonumber \\
&=&{c\over a}\Gamma\left({i\Om c\over a}\right)\left({\om_Kc\over a}\right)^{-i\Om c/a}
e^{-\pi\Om c/2a} \ ,
\label{eq7}
\eea
where $\Gamma$ is the gamma function\cite{gr}. Then, since \cite{abramowitz} $\left|\Gamma\left({i\Om c\over a}\right)\right|^2=
\pi/[(\Om c/a)\sinh(\pi\Om c/a)]$,
\be
\left|\int_{-\infty}^{\infty}d\tau e^{i\Om\tau}e^{i(\om_Kc/a)e^{a\tau/c}}\right|^2 =
{2\pi c\over \Om a}{1\over e^{2\pi\Om c/a}-1}.
\label{eq8}
\ee
The time-dependent Doppler shift seen by the accelerated observer therefore
leads to the Planck factor $(e^{\hbar\Om/kT}-1)^{-1}$
with $T=\hbar a/2\pi kc$, which is just equation \Eq{eq1}. We obtain the same result in the
case of a wave propagating in the $+{\bf z}$ direction.

Note that the time-dependent phase can also be obtained directly by considering the standard nonaccelerated
Minkowski plane wave $\exp[i \varphi_\pm] \equiv \exp[i(K z \pm \om_K t)]$ and using equations \Eq{eq3.5}:
$\varphi_\pm(\tau) = K z(\tau) \pm \om_K t(\tau) = (\om_K c/a) \exp(\pm a \tau /c)$
with $K = \om_K/c\,$ \cite{pringle,kolbenstvedt}.

\section{A More Formal Derivation}\label{formal_derivation}
The ``derivation" of the temperature \Eq{eq1} just given leaves much to be desired. We have restricted ourselves
to a single field frequency $\om_K$, whereas a quantum field in vacuum has components at all frequencies. Moreover
we have noted the appearance of the Planck factor but have not actually compared our result to that appropriate to
an observer at rest in a {\it thermal} field.

To rectify these deficiencies, let us consider a massless scalar field in one spatial dimension ($z$), quantized in a box
of volume $V$ \cite{remark}:
\be
\hat{\phi}=\sum_K\left({2\pi\hbar c^2\over\om_KV}\right)^{1/2}\left[\hat{a}_Ke^{-i\om_Kt}+\hat{a}_K^{\dag}e^{i\om_Kt}\right] \ .
\label{eqa1}
\ee
Here $K=\pm\om_K/c$, and $\hat{a}_K$ and $\hat{a}_K^{\dag}$ are respectively the annihilation and creation operators
for mode $K$ ($[\hat{a}_K,a_{K'}^{\dag}]=\delta_{KK'}$,
$[\hat{a}_K,a_{K'}]=0$). We use a caret ( ${\hat{}}$ ) to denote quantum-mechanical  operators.
The expectation value $\langle(d\hat{\phi}/dt)^2\rangle/4\pi c^2$ of the
energy density of this field is
 $V^{-1}\sum_K\hbar\om_K[\langle \hat{a}_K^{\dag}\hat{a}_K\rangle +1/2]$. For simplicity we consider the field at a particular point in space (say, $z=0$),
since spatial variations of the field will be of no consequence for our purposes.

For a thermal state the number operator $\hat{a}_K^{\dag}\hat{a}_K$ has the expectation
value $(e^{\hbar\om_K/kT}-1)^{-1}$. Consider the Fourier transform operator
\be
\hat{g}(\Om)={1\over 2\pi}\int_{-\infty}^{\infty}dt\hat{\phi} e^{i\Om t}=\sum_K\left({2\pi\hbar c^2\over \om_KV}\right)^{1/2}\hat{a}_K\delta(\om_K-\Om) , \ \ \Om >0.
\label{eqa2}
\ee
The expectation value $\langle \hat{g}^{\dag}(\Om)\hat{g}(\Om')\rangle$ in thermal equilibrium is therefore
\bea
\langle \hat{g}^{\dag}(\Om)\hat{g}(\Om')\rangle&=&\sum_K\left({2\pi\hbar c^2\over\om_KV}\right)\langle \hat{a}_K^{\dag}\hat{a}_K\rangle
\delta(\Om-\Om')\delta(\om_K-\Om) \nonumber \\
&=&\sum_K\left({2\pi\hbar c^2\over\om_KV}\right){1\over e^{\hbar\om_K/kT}-1}
\delta(\Om-\Om')\delta(\om_K-\Om) \ .
\eea
We go to the limit where the volume of our quantization box becomes very large, $V\rightarrow\infty$,
so that we can replace in the usual fashion the sum over $K$ by an integral:
$\sum_K\rightarrow (V/2\pi)\int dK$ \cite{remark1}. Thus
\bea
\langle \hat{g}^{\dag}(\Om)\hat{g}(\Om')\rangle&=&\hbar c^2\int_{-\infty}^{\infty}dK{1\over\om_K}{1\over
e^{\hbar\Om/kT}-1}\delta(|K|c-\Om)\delta(\Om-\Om') \nonumber \\
\mbox{}&=&{2\hbar c/\Om\over e^{\hbar\Om/kT}-1}\delta(\Om-\Om') \ .
\label{thermal}
\eea

Now let us consider an observer in uniform acceleration in the quantized {\it vacuum} field. This observer sees each field frequency
Doppler-shifted according to \Eq{eq4} and \Eq{eq5}, and so for him the operator $\hat{g}(\Om)$ has the form
\bea
\hat{g}(\Om)&=&{1\over 2\pi}\int_{-\infty}^{\infty}d\tau e^{i\Om\tau}\sum_K\left({2\pi\hbar c^2\over\om_KV}\right)^{1/2}\left[\hat{a}_K
e^{-i\int^{\tau}d\tau'\om_K'(\tau')}+\hat{a}_K^{\dag}e^{i\int^{\tau}d\tau'\om_K(\tau')}\right] \nonumber \\
&=&{1\over 2\pi}\int_{-\infty}^{\infty}d\tau e^{i\Om\tau}\sum_K\left({2\pi\hbar c^2\over\om_KV}\right)^{1/2}[\hat{a}_K
e^{i(\epsilon_K\om_Kc/a)e^{-\epsilon_Ka\tau/c}} \nonumber \\
&\mbox{}& + \hat{a}_K^{\dag}e^{-i(\epsilon_K\om_Kc/a)e^{-\epsilon_K\om_K\tau/c}}],
\label{eq10}
\eea
where $\epsilon_K=|K|/K$. Since $\hat{a}_K|{\rm vacuum}\rangle=0$, only the $\hat{a}_K^{\dag}$
terms in this expression contribute to the vacuum expectation
value $\langle \hat{g}^{\dag}(\Om)\hat{g}(\Om')\rangle$. Performing the integrals over $\tau$ as before, and using
$\langle \hat{a}_Ka_{K'}^{\dag}\rangle=\delta_{KK'}$,
 we obtain
\be
\langle \hat{g}^{\dag}(\Om)\hat{g}(\Om')\rangle=\left({c\over 2\pi a}\right)^2\left({2\pi\hbar c^2\over V}\right)\left|\Gamma\left({i\Om c\over a}\right)\right|^2
e^{-\pi c\Om/a}\sum_K{1\over\om_K}\left({\om_Kc\over a}\right)^{i\epsilon_K(\Om-\Om')c/a} \ ,
\label{eq12}
\ee
where we use the fact that the sum over $k$ vanishes unless $\Om-\Om'$. In fact we show in the Appendix that the sum over $K$ is
$(2Va/c^2)\delta(\Om-\Om')$, so that
\be
\langle \hat{g}^{\dag}(\Om)\hat{g}(\Om')\rangle={\hbar c^2\over\pi a}\left|\Gamma\left({i\Om c\over a}\right)\right|^2e^{-\pi c\Om/a}\delta(\Om-\Om')
={2\hbar c/\Om\over e^{2\pi\Om c/a}-1}\delta(\Om-\Om') \ ,
\label{eq13}
\ee
which is identical to the thermal result \Eq{thermal} if we define the temperature
by equation \Eq{eq1}.

Note that the expectation value  $\langle \hat{a}_Ka_{K'}^{\dag}\rangle=\delta_{KK'}$
involves the creation and annihilation operators of the accelerated observer
and is taken with respect to the
{\it accelerated observer's} vacuum, which is different from the vacuum seen by the nonaccelerated observer.
This point is discussed more fully in  Section~\ref{discussion}.

\section{Fermi-Dirac Statistics for Dirac Particles}
In the above we have considered a scalar field and have derived the
Planck factor $(e^{\hbar\Om/kT}-1)^{-1}$ indicative of Bose-Einstein (BE) statistics.
We began with the standard plane-wave solutions of the form $\exp[i(K z \pm \om_K t)]$ for the nonaccelerated
Minkowski observer, and considered the time-dependent Doppler shifts as seen by the accelerated observer.
For the spin-$1/2$ Dirac particles one would expect an analogous derivation to reproduce the
Planck factor $(e^{\hbar\Om/kT}+1)^{-1}$ indicative of Fermi-Dirac (FD) statistics.

We now show that this is indeed
the case. Mathematically, the essential point involves the replacement
$i\Om c/a \to i\Om c/a + 1/2$ in the integrals in \Eq{eq6}- \Eq{eq8} \cite{dirac_note},
and the relationship $|\Gamma(i\Om c/a + 1/2)|^2 = \pi/\cosh(\pi \Om c/a)$ \cite{abramowitz}.
Physically, this replacment arises from
the additional spinor nature of the Dirac wave function over that of the scalar plane wave. In the case of a scalar
field, only the phase had to be instantaneously Lorentz-transformed to the comoving frame of
the accelerated observer. For non-zero spin, the spinor structure of the particles must also
be transformed \cite{weinberg}, or "Fermi-Walker transported" \cite{fermiwalker} along a particle's trajectory
to ensure that it does not ``rotate" as it travels along the accelerated trajectory. This leads to
a time-dependent Lorentz transformation of the Dirac bispinor of the form \cite{S_transform}
$\hat{S}(\tau) = \exp(\gamma^0 \,\gamma^3\, a\tau/2c) = \cosh(a\tau/2c) + \gamma^0 \,\gamma^3\,\sinh(a\tau/2c)$,
where the $4\times 4$ constant Dirac matrices are given by
$$
\gamma^0 = \left(%
\begin{array}{cc}
  1 & 0 \\
  0 & -1 \\
\end{array}%
\right), \qquad
\gamma^3 = \left(%
\begin{array}{cc}
  0 & \sigma_z \\
  -\sigma_z & 0 \\
\end{array}%
\right),
$$
and $\sigma_z = {\rm diagonal}(1,-1)$
is the usual $2\times 2$ Pauli spin matrix in the $\mathbf{z}$ direction.
Acting on a spin up state $\mid\uparrow\rangle = [ 1, 0, 1, 0]^T$ \cite{greiner}, $\hat{S}(\tau)$ gives
$\hat{S}(\tau) \mid\uparrow\rangle = \exp(a\tau/2c)\,\mid\uparrow\rangle$. Thus, for the spin up
Dirac particle we should replace the plane wave scalar ``wave function"  $\exp[i\varphi(\tau)]$ used in
\Eq{eq6} with $\exp(a\tau/2c) \, \exp[i\varphi(\tau)]$ \cite{haycan}.
This leads to the replacement $i\Om c/a \to i\Om c/a + 1/2$ in \Eq{eq7}, and therefore the result
\be
\left|\int_{-\infty}^{\infty}d\tau e^{i\Om\tau}\,e^{a\tau/2c}\,e^{i(\om_Kc/a)e^{a\tau/c}}\right|^2 =
{2\pi c\over \om_K a}{1\over e^{2\pi\Om c/a}+1} \ .
\label{FDspectrum}
\ee

Comparing \Eq{FDspectrum} with \Eq{eq8} we note the change of sign in the denominator from
$-1$ for BE statistics to $+1$ for FD statistics. We also note that the prefactor in \Eq{eq8}
involves the dimensionless frequency $\Om c /a$ while in \Eq{FDspectrum} the prefactor involves
the factor $\om_K c /a$ (the argument of the exponential in the distribution function is still $\hbar\Om/kT$ with
the same Unruh temperature $T = \hbar a/ 2\pi k c$).
This is no cause for concern, since in fact a single Minkowski frequency $\om_K$ is actually
spread over a continuous range of accelerated (Rindler) frequencies $\Om$ with peak centered
at $\Om = \om_K$ \cite{freqspread}. This allows us to replace $\om_K$ by $\Om$ in the final result.
(This is explicitly evidenced by the delta function $\delta(\om_K - \Om)$ in \Eq{eqa2} - \Eq{eq13} in the
comparison of the thermal and accelerated correlation functions.)

For the case of the spin up Dirac particle, the more formal field-theoretic derivation of Section \ref{formal_derivation}
proceeds in exactly the same fashion, with the modification of the accelerated wave function from
$\exp[i\varphi(\tau)] \to \exp(a\tau/2c) \, \exp[i\varphi(\tau)]$ and the use of
anti-commutators $\{\hat{a}_K,a_{K'}^{\dag}\}=\delta_{KK'}$ for the quantum-mechanical creation and annihilation
operators instead of the commutators appropriate for scalar BE particles. For the correlation
function we find
\be
\langle \hat{g}^{\dag}(\Om)\hat{g}(\Om')\rangle
={2\hbar c/\om_K\over e^{2\pi\Om c/a}+1}\delta(\Om-\Om') \ ,
\label{Dirac_correl}
\ee
the FD analogue of \Eq{eq13}.

\section{Summary and Discussion}\label{discussion}
In the usual derivation of the Unruh temperature \Eq{eq1}, \cite{unruh,davies,also,takagi,davies2}
one solves the wave (or Dirac) equation for the field mode functions in the Rindler coordinates \Eq{eq3.5}
and then quantizes them.
Since the hyperbolic orbit of the accelerated observer is confined to the (right Rindler) wedge $z > |t|$ (with "mirror"
orbits in the left Rindler wedge $z < -|t|$) it turns out that the vacuum seen by the accelerated observer
in say, the right Rindler wedge is different than the usual Minkowski vacuum (defined for all $z$ and $t$)
seen by the unaccelerated observer. The inequivalence of these vacua
(and hence the Minkowski vs. Rindler quantization procedures \cite{fulling}) is due to the fact that the
right and left Rindler wedges are causally disconnected from each other. This can be easily seen by drawing a Minkowski
diagram in $(z,t)$ coordinates and observing that light rays at $\pm 45^{\circ}$ emanating from one wedge do not penetrate the other wedge.
Hence the Minkowski vacuum that the accelerated observer moves through appears to him as an excited state containing particles,
and not as the vacuum appropriate for the right Rindler wedge.
The Bose-Einstein distribution with Unruh temperature $T$ for scalar fields (Fermi-Dirac for Dirac fields)
then arises by considering the expectation value of the number operator $\hat{a}_R^\dagger \hat{a}_R$ for the \textit{accelerated} observer
(in the right Rindler wedge) in the \textit{unaccelerated} Minkowski vacuum $| 0_M\rangle$, i.e.
$\langle 0_M | \hat{a}^\dagger_R \hat{a}_R| 0_M\rangle \sim (\exp(\hbar\Omega/kT)\pm 1)^{-1}$ (with upper sign for scalar fields
and lower sign for Dirac fields).
This is referred to as the \textit{Thermalization Theorem} by Takagi \cite{takagi}.

In this work we take a slightly different viewpoint \cite{pwm_book}.
For a scalar field, we first consider an unaccelerated Minkowski observer in a \textit{thermal} state and find that
the expectation value of a field correlation function is proportional to the Bose-Einstein distribution.
We then consider the calculation of this correlation function again, but this time for
an accelerated Rindler observer in his \textit{Rindler vacuum}\cite{vacua} state $| 0_R\rangle$ , such that for
a single mode, $\langle 0_R | \hat{a}_R \hat{a}^\dagger_R | 0_R\rangle = 1$.
The new feature now is that from his local stationary perspective, the accelerated observer perceives all Minkowski frequencies (arising from
the the usual plane waves associated with Minkowski states) as time-dependent Doppler shifted frequencies.

The derivation presented in this work shows why quantum field fluctuations in the vacuum state are crucial for the thermal effect of
acceleration: $\langle \hat{g}^{\dag}(\Om)\hat{g}(\Om')\rangle$ is nonvanishing because the vacuum expectation
 $\langle \hat{a}_K\hat{a}_K^{\dag}\rangle\ne 0$.
But there's more to it than that, because $\langle \hat{a}_K\hat{a}_K^{\dag}\rangle$ is also
nonvanishing for an observer with $a=0$. For such an
observer, however,
\be
\int_{-\infty}^{\infty}d\tau e^{i\Om\tau}e^{i\int^{\tau}d\tau'\om_K(\tau')}=
\int_{-\infty}^{\infty}d\tau e^{i(\Om+\om_K)\tau}=0 \
\label{disc1}
\ee
for the case of scalar particles.
In other words, the thermal effect of acceleration in our model arises because of the
nontrivial nature of the quantum vacuum {\it and} the
time-dependent Doppler shifts seen by the accelerated observer.
For the case of Dirac particles, the essential new feature is the additional spinor structure of the
wave function over that of the scalar plane wave. In order to keep the spin "non-rotating" in
the comoving frame of the accelerated observer, the Dirac bispinor must be Fermi-Walker transported
along the accelerated trajectory, resulting in an additional time-dependent Lorentz transformation.
Formally, this induces a shifting of $i\Om c/a \to i\Om c/a + 1/2$ in the calculation of relevant gamma function-like integrals, leading
to the FD Planck factor.

In the following we briefly discuss the relationship of our correlation function to those
used in the standard literature on this subject and point out a not widely appreciated curious subtlety relating
details of the spatial Rindler mode functions (which we have ignored in our model) to the statistics of
the noise spectrum seen by the accelerated observer.

In our model, we have not given any motivation for using the correlation function $\langle \hat{g}^{\dag}(\Om)\hat{g}(\Om')\rangle$ aside from the
fact that we could calculate it for a nonaccelerated observer in a thermal field and for a uniformly accelerated observer in vacuum and
compare the results. It is easy to show that a harmonic oscillator with frequency $\om_0$ and dissipation coefficient
$\gamma$, linearly coupled to the field \Eq{eqa1}, reaches a steady-state energy expectation value
\be
\langle E\rangle\propto \int_0^{\infty}d\Om\int_0^{\infty}d\Om'{\langle \hat{g}^{\dag}(\Om)\hat{g}(\Om')\rangle\over
(\Om-\om_0-i\gamma)(\Om'-\om_0+i\gamma)} \ ,
\ee
which offers some motivation for considering $\langle \hat{g}^{\dag}(\Om)\hat{g}(\Om')\rangle$.
In fact, it can be shown that $\langle E\rangle =[e^{\hbar\om_0/kT}-1]^{-1}$,
which shows again that our accelerated observer acquires  the characteristics
appropriate to his being in a thermal field at the temperature $T=\hbar a/2\pi kc$.

In an extensive review of the Unruh effect, Takagi \cite{takagi} (Chapter 4) utilizes the quantum two-point
correlation (Wightman) function
$g_W(\tau,\tau') \equiv \langle 0_M\mid \hat{\phi}(\tau) \,\hat{\phi}^\dagger(\tau')\mid 0_M\rangle$
to determine the power spectrum of the vacuum noise seen by the accelerated observer
for a scalar field via
$$
S(\Om) \equiv \lim_{s\downarrow 0} \int^\infty_\infty e^{-i\Om\tau - s\mid\tau\mid} g_W(\tau),
$$
which is very much in the spirit of our calculation.
Here the field operator
$\hat{\phi}(\tau)$ is expanded in terms of the Rindler mode functions and involve
creation and annihilation operators for both the right and left Rindler wedges.
Takagi shows the remarkable, though not widely known result, that for a scalar field in
a Rindler spacetime of dimension $n$,
$S_n(\Om) \sim f_n(\Om)/[\exp(\hbar\Om/kT) - (-1)^n]$.
For even-dimensional spacetimes (e.g. $n=2$ considered in this work, or the usual $n=4$)
$S_n(\Om)$ is proportional to the Bose-Einstein distribution function $[\exp(\hbar\Om/kT) - 1]^{-1}$, and essentially reproduces
our equation \Eq{eq13} (up to powers of $\Om c /a$, contained in the function $f_n(\Om)$).
However, for $n$ odd, $S_n(\Om)$ is actually
proportional to the Fermi-Dirac distribution $[\exp(\hbar\Om/kT) + 1]^{-1}$.
For the case of Dirac particles
the opposite is true, namely for even spacetime dimensions $S_n(\Om)$ is proportional to the FD distribution
and for odd spacetime dimensions $S_n(\Om)$ is proportional to the BE distribution. This curious fact
arises from the dependence of $S_n(\Om)$ on two factors in its calculation. The first is
the above mentioned \textit{Thermalization Theorem}, namely that
the number spectrum of accelerated (Rindler) particles in the usual nonaccelerated Minkowski vacuum
is proportional to the BE distribution function for scalar fields and is proportional to the FD distribution function
for Dirac fields.
The second factor that switches the form of $S_n(\Om)$ from BE to FD depends on the detailed
form of the Rindler mode functions (see Takagi\cite{takagi}, Chapter 4 for more complete details).
Though the trajectory of the accelerated observer takes place in ``1+1" dimensions (i.e. the $(z,t)$ plane), the quantum field exists
in the full $n$-dimensional spacetime, and thus $S_n(\Om)$ ultimately depends of the form of the mode functions
in the full spacetime. In spacetimes of even dimensions the number spectrum of Rindler particles
in the Minkowski vacuum and the noise spectrum of the vacuum fluctuations (i.e., the response of the
accelerated ``particle detector") both depend on the same distribution function, and these two effects
are often incorrectly equated.

In our simplified derivation we have bypassed this technicality by performing our calculations in
``1+1" dimensions (i.e., $n=2$). We have concentrated on the power spectrum of vacuum fluctuations
as seen by a particle detector carried by the accelerated observer. We have shown that
in ``1+1" dimensions the spectrum of fluctuations is proportional to the Bose-Einstein distribution
function for scalar fields and to the Fermi-Dirac distribution for spin-1/2 fields, with
the Hawking-Unruh temperature defined by \Eq{eq1}.
The dependence of the noise spectrum on these distribution functions is ultimately traced back
to the time-dependent Doppler shifts as seen by the accelerated observer as he moves through
the usual nonaccelerated Minkowski vacuum. It is hoped that the simple calculations exhibited here are
straightforward enough to give an intuitive understanding of the
essential physical origin of the Hawking-Unruh temperature experienced by a uniformly
accelerated observer.

\flushleft{\bf Exercise for the student}
Discuss when the Hawking-Unruh temperature from \Eq{eq1} would become physically detectable by
utilizing the expression $a=G M/r^2$ for the gravitational acceleration of a test mass at a distance $r$ from
a mass $M$, and determining $T$   at the surface of the earth, the sun and
a Schwarzschild black hole.

\section*{Appendix}
Converting the sum over $K$ to an integral, we have, for $K>0$,
\be
\sum_{K>0}{1\over\om_K}\left({\om_Kc\over a}\right)^{i\epsilon_K(\Om-\Om')c/a}={V\over 2\pi}\int_0^{\infty}dK{1\over\om_K}
\left({\om_Kc\over a}\right)^{i(\Om-\Om')c/a} \ .
\ee
Letting $x=\log(\om_Kc/a)$, we can write this as
\be
{V\over 2\pi c}\int_{-\infty}^{\infty}dxe^{-ix(\Om-\Om')c/a}={Va\over c^2}\delta(\Om-\Om') \ .
\ee
The same result is obtained for the sum over $K<0$, so that the sum over all $K$ is $(2Va/c^2)\delta(\Om-\Om')$.

\section*{Acknowledgements}

Thanks go to H. Fearn, D. McMahon, G. J. Milburn  E. Mottola, M.
O. Scully  and M. Wolinsky for brief discussions relating to
acceleration in vacuum.


\begin{references}
\bibitem{remark}  When we later convert a sum over $K$ to an integral, we obtain the Lorentz-invariant measure $dK/\om_K$ as
a consequence of the $1/\sqrt{\om_K}$ in \Eq{eqa1}. This eliminates the need to explicitly
transform this quantity in \Eq{eq10}, for
instance.
\bibitem{remark1} Since we use a one-dimensional model the factor $V/2\pi$ appears instead of the more familiar $V/(2\pi)^3$. In
other works, our volume $V$ here is really just a length.
\bibitem{dirac_note} Using the same gamma function integrals as in Reference [8] above, we will now have
for the Dirac case $\mu = i\Omega c/a + 1/2$, with $Re(\mu)=1/2$ clearly in their domain of definition $0<Re(\mu)<1$.
\bibitem{haycan} The spinor Lorentz transformation $\hat{S}(\tau)\,$ does not mix spin components. Thus, for e.g.,
a spin up Minkowski state remains a spin up accelerated (Rindler) state. We can therefore drop the
constant spinor $\mid \uparrow \rangle$ from our calculations and retain the essential, new time-dependent
modification $\exp(a\tau/2c)$ to the plane wave for our Dirac "wave function."
\end{references}
\end{document}